\documentclass[]{article}
\usepackage[english,activeacute]{babel}
\usepackage{graphicx}
\usepackage{amsmath,amssymb} 
\usepackage{xspace}
\usepackage{color}
\usepackage{pdfsync}
\usepackage{algorithm}
\usepackage{algorithmicx}
\usepackage[percent]{overpic}
\algrenewcommand\alglinenumber[1]{\small #1:}
\graphicspath{figures}

\begin{document}

\title{Using emulation to validate applications on opportunistic networks}

\author{Gwilherm Baudic, Antoine Auger, Victor Ramiro, Emmanuel Lochin\\
\texttt{\{first.second@isae.fr\}}\\
Institut Sup\'{e}rieur de l'A\'{e}ronautique et de l'Espace\\
ISAE-SUPAERO, Universit\'{e} de Toulouse\\
}
\date{}

\maketitle
\begin{abstract} 
The increasing trend on wireless-connected devices makes opportunistic networking a promising alternative to existing infrastructure-based networks. However, on these networks there is neither guarantee about the availability of the connections nor on the topology of the network. The development of \textit{opportunistic applications}, i.e., applications running over opportunistic networks, is still in early stages. This is due to lack of tools to support the process in such uncertain conditions. Indeed, many tools have been introduced to study and characterize opportunistic networks, but none of them is focused on helping developers to conceive opportunistic applications. In this paper, we argue that the gap between opportunistic applications development and network characterization can be filled with network emulation. First, this position paper points out important challenges about the development of opportunistic applications. Then, in order to cope with these challenges, it details a set of requirements that an emulator should meet to allow the testing of such applications.
\end{abstract}

%!TEX root = main.tex

\section{Introduction}
\label{sec:introduction}

%Delay tolerant networks (DTNs) were introduced to deal with environments where interruptions or disruptions of service were expected. Such networks usually lack end-to-end paths or any infrastructure to help communications. In its purest form, the definition of a DTN is based on the delay tolerance nature of communications; it covers a wide range of networks: space communications, vehicular networks, sensor networks, opportunistic networks, etc. We can associate a DTN to the general case of a network which may evolve with some unknown underlying process. Usually, there is neither guarantee about the availability of the connections nor the topology of the network.

Opportunistic networks are a special case of DTNs~\cite{kevin_fall_DTN} where nodes systematically exploit their mobility to benefit from contacts as a communication opportunity to forward messages. This mobility introduces delays when a node cannot forward its message, keeping it in its own buffer. This allows routing protocols to exploit opportunistic contacts, in absence of stable end-to-end path, as a means to create a temporal path for delivery. The store-carry-forward paradigm allows nodes to exploit spatiotemporal paths created by contact opportunities in order to deliver messages over time. Opportunistic networks are also suitable for communications in pervasive environments that are saturated by other devices. The ability to self-organize using the local interactions among nodes, added to mobility, leads to a shift from legacy packet-based communications towards a message-based communication paradigm.

However, dealing with the dynamics of opportunistic networks is complicated~\cite{conti2014}. Let us consider the number of parties involved in the network, for instance $n$ inhabitants in a medium dense city. First, the number of interactions between them grows as $O(n^2)$. Secondly, the continuously changing topology, due to the nodes mobility and its interactions, leads to an explosion of the number of states needed to characterize the behavior for any algorithm to be deployed. The number and density of nodes, their interactions, their mobility, the different routing strategies impact on the delay, packet lost, retransmissions, etc. Different ways have been proposed to study those dynamics. The main focus of research up to now was to define an optimal routing strategy to deliver some domain specific information, but they do not consider the final application development.

With the current trend on connected devices, the idea of \textit{opportunistic applications}, i.e., applications running over opportunistic networks, is getting closer to being a reality. However, there are still several obstacles remaining before we see a massive deployment over this paradigm. Indeed, to conceive applications working on these networks remains one of the biggest challenges to overcome. This can be explained by the complexity to undertake a performance evaluation of an opportunistic application before a real world deployment.

% talk about other approaches
Simulations offer a fast and lightweight way of getting insight in the behavior of the network~\cite{opnet, ns3, the_one}. Unfortunately, they do not give a simple way of thinking in terms of real applications, and typically cannot derive metrics related to QoE because they usually focus on purely network-related performance. Testbeds~\cite{beuran_dtn_2013,beuran_making_2012,
li_tunie_2015,giordano_movit:_2012,mobility_nicta,
morgenroth_hydra:_2010,bittencourt_towards_2013,zhou_twine:_2006,maeda_hybrid_2008,
liu_open_2007,zhang_design_2009} are effective in terms of an almost real world feedback, but they are really expensive to deploy in the middle of the development process.

Developers of opportunistic applications must not only deal with network characterization, but also with its impact on the application. Filling this gap between network characterization and development should be supported by current developments tools. However, even the ability to test the applications conformity to a simple DTN messaging protocol is missing today. \textit{We need to better integrate how developers consider network metrics obtained from the characterization phase into the development process of opportunistic applications}. 

%ABSTRACT
%First, this position paper points out important challenges about the development of opportunistic applications. Then, in order to cope with these challenges, it details a set of requirements that an emulator should meet to allow the testing of such applications.

%In this work, we focus on helping developers to conceive their opportunistic applications. 
%and we propose a new hybrid emulation system for opportunistic networks.
%we address the limitations of current approaches, highlighting the gap between network characterization and development. 

The rest of the paper is structured as follows. We first introduce the challenges related with opportunistic networks in Section~\ref{sec:challenges}. It allows us to highlight the gap between network characterization and application development. In order to cope with this challenge, we define important requirements, presented in Section~\ref{sec:requirements}. Finally, we conclude and give some directions for future work in Section~\ref{sec:conclusions}.

%!TEX root = main.tex

\section{Development Challenges of Opportunistic Applications}
\label{sec:challenges}

In this section, we discuss the challenges when developing opportunistic applications. We see those challenges from two perspectives: the way developers deal with the network characterization and the way they assess the network impact on the application. Finally, we highlight the existence of a gap between these two perspectives.

\subsection{Dealing with Opportunistic Networks Characterization}

As we said, dealing with the dynamics of opportunistic networks is complicated. In this section, we discuss the main challenges when dealing with opportunistic networks from a developer's point of view. We discuss the current available alternatives (and their drawbacks) to characterize opportunistic networks.

\subsubsection{Analytical modeling}

Several analytical models have been presented to characterize opportunistic networks. The main goal of analytical models is to provide a closed formula for a specific characteristics. Others propose algorithms to approach reality~\cite{groenevelt, aoc2015, haas_model, zhang_model}. However, most of these models assume unrealistic hypotheses or they cannot scale. Indeed, the number of states needed to model DTNs increases with the interactions parties have in the network. The number of interactions and the changes they follow makes this problem a highly combinatorial one. Most of them belong to the NP-class. 

\subsubsection{DTN simulators}

Opportunistic networks simulators~\cite{the_one, opnet, ns3} are mainly focused on nodes mobility and efficient routing issues. Indeed, most of the research on DTNs has been focused on the message routing problem as the application. 

On the one hand, we find many tailored simulators for specific cases. On the other hand, we find an effort to standardize the results based on The~ONE~\cite{the_one} simulator. Nevertheless, current simulators do not allow developers to plug real devices to interact with them. For instance, The~ONE simulator provides a simulated network stack. This stack includes a network interface layer, a connection layer (constant and variable bit-rates), a routing layer (with several routing algorithms) and, finally, an application layer on the top. However, this application layer is just a basic handler class for messages passed by from the simulated routing protocol. For a real developer, the impossibility to think in terms of a concrete user application, independently of all those complexities, is still a huge problem.

\subsubsection{Traces collection}

Another effort to better understand opportunistic networks has been the development of devices and applications to collect peer contact traces. The characterization of the contact and intercontact time distribution allows to understand the dynamics behind the network. Ideally, this abstraction in terms of contact should be independent from the link layer. However, this is not the case in reality~\cite{gb-chants}. Indeed, current communication layers lack the characteristics needed to really deploy opportunistic applications. This makes the collected traces to be less representative than we need, and therefore hard to replay in order to help application developers.  

\subsubsection{DTN emulators and testbeds}

Emulation naturally provides a bridge between simulation and real world testing~\cite{emulation-book}. It consists of putting together real and simulated components in a single system. In existing proposals, real parts are most of the time the application and underlying operating system, while the network is simulated. For instance, KauNet~\cite{garcia2007kaunet} is a pattern-based link emulator for mobile and wireless systems. In~\cite{tanguy}, P\'{e}rennou et al. extend it to support opportunistic networks, thanks to trigger mechanisms.

More generally, emulation is seen as a convenient way to get closer to the realism of field trials, while at the same time offering the repeatability and scalability of simulations. A good emulation system must also be completely transparent to the real part and should offer sufficient flexibility to be used with various mobility models. However, scalability is often achieved by using testbeds, which are hard and costly to setup.

\subsubsection{Middlewares and Real DTN stacks}

Most of DTN middlewares (e.g., MaDMAN~\cite{petz_madman:_2010} or the solution of Jiang et al.~\cite{jiang_publish/subscribe_2011}) focus on integration and architectural considerations, arguing that human mobility will address connectivity issues and that applications will keep working correctly within a DTN context. Although these solutions can be considered as proofs of concept, they do not provide any useful tools nor metrics to potential applications developers. Plus, the proper functioning of these middlewares is only assessed on the field, by looking at the number of packets delivered. Finally, the results are only valid for a given application and for a given context.

Implementations of the DTN stack are still in early stage. We count among them ION~\cite{ION}, DTN2~\cite{DTN2} or IBR-DTN~\cite{IBR-DTN}. Even though they exist, they are not yet massively deployed in any final user platform. 

% To delete ?
%Middlewares are also a way to interact with opportunistic networks. For instance, MaDMAN \cite{petz_madman:_2010} is a middleware for delay-tolerant mobile ad-hoc networks. Since it continuously analyzes Context information, it is able to switch to the best suitable network stack (DTN or TCP) at a given time, preserving applications' communication sessions (and more specifically TCP acknowledgements). 

% To delete ?
%In \cite{jiang_publish/subscribe_2011}, Jiang et al. present a mobile middleware for resilient communications. This middleware combines both Publish/Subscribe message-oriented middleware (MOM) and DTN paradigms. Bundle messages can be sent using either TCP or Bluetooth network stacks. When many are available, network selection is made on network availability, user preference, and energy left. This distributed middleware has been deployed on Android phones to ensure resilient communications in a neighborhood watch system use case.

% We strongly believe that DTN emulation can help developers to take right decisions to dimension and test the proper functioning of their applications before deploying them. By providing advanced real-time monitoring features, our DTN emulator can be used as a convenient testing tool by developers.

% Victor
% tu veux aussi parler de cette these?
% https://www.researchgate.net/publication/272148501_RON_-_Opportunistic_Networks
% \cite{RONOpportunisticNetworkG}

\newpage
\subsection{Dealing with Opportunistic Networks Impact}

Within an opportunistic network context, end-to-end delays, delivery ratio and drop ratio are very important factors that developers want to study before deploying their applications. However, this set of network-related metrics is not sufficient to correctly describe the impact of those networks for the end user.

Quality of Experience (QoE) is defined by the \mbox{ITU-T} as \textit{``the overall acceptability of an application or service, as perceived subjectively by the end-user"}~\cite{itu-t_g.1080_2008}. As a matter of fact, QoE has objective and subjective dimensions. Although subjective dimension refers to human components (such as emotions, ease of use, etc.), objective dimension relates to more measurable and quantifiable factors (such as network parameters for instance). Developers need to test their applications from a user viewpoint (subjective dimension) while having an overview of some key metrics in real time (objective dimension). 

In line with the vision of Brooks et al.~\cite{brooks_user_2010}, we believe that it is relevant to express QoE as a combination of user experience and technical measurements. In an emulator, we assume that technical measurements should refer to network-related metrics, while user experience relies on the following question: \textit{Is my application working as I expected?} Please note that several measures of user experience exist and can also be envisioned, including for mobile use cases~\cite{park_modeling_2013}.

%Based on this QoE model, our emulator provides more accurate insights into the minimum network characteristics that an application requires. By using our emulator, developers can now figure out what is changed in the use of their application within a DTN context, as well as the set of minimum requirements that a network should provide.

\subsection{Reconciling Perspectives}

We argue there is a gap between opportunistic network characterization and applications development.

On the one hand, network characterization provides metrics to better understand the underlying behavior of the opportunistic network. Indeed, opportunistic and DTN networks have lead to a change from a connection-oriented to a message-based paradigm. Opportunistic applications rely on the store-and-forward paradigm to correctly work. Some properties of these networks, such as the non guarantee of end-to-end paths, make impossible to ignore the characterization phase when developing opportunistic applications. 

On the other hand, developers may not have the adequate knowledge or resources to take advantage of the network characterization. Instead, they want to know, possibly in a quick and simple way, if \emph{their} applications will \emph{still} work within an opportunistic use case. Making hypotheses on the underlying network (end-to-end delay, drop ratio, etc.) often leads to over provisioning and resources waste.

Hence, there is a gap in the way developers deal with network metrics obtained from network characterization. In the next section, we motivate the need to use emulation to reconcile both.

%!TEX root = main.tex

\section{Requirements for Opportunistic Network Emulation} % (fold)
\label{sec:requirements}

As presented in Section~\ref{sec:challenges}, there is a gap between network characterization and application development. We argue that an opportunistic network emulator can fill this gap. In the following, we distill the main requirements for opportunistic network emulation. Emulation provides an environment to deploy real applications, helping developers to craft opportunistic applications. 

\subsection{Link Layer Requirements}

One problem of opportunistic networks is the lack of a clear communication technology that can deal with the evaluation of real applications. Indeed, current link layer technologies, such as Bluetooth, Wi-Fi~Direct or other ad-hoc communication technologies lack many of the assumptions made by opportunistic networks algorithms. 

Nevertheless, we do not see this as a real problem but rather as an opportunity. Indeed, based on the idea of traces collection, we propose to focus on the contact and intercontact time characteristics of the network over a proper link layer. Contacts between nodes can easily be emulated by adding some local parameters to represent delays or packet losses on a pair-wise basis, allowing in some sort a worst case scenario analysis. Hence, we define two requirements that we assume as hypotheses to be met by any communication technology to support opportunistic applications:

\begin{itemize}
	\item[C1] \textbf{Connection bi-directionality:} Algorithms proposed over opportunistic networks can assume connections to be bi-directional. Consider for example the summary vector exchange in epidemic routing, which is impossible if the connection is not bi-directional.
	\item[C2] \textbf{Multicast communication:} Nodes can connect with multiple devices at the same instant of time over opportunistic networks. %source?
\end{itemize}

\subsection{Connection-oriented vs Contact-oriented emulation}

Simulators are usually implemented in a connection-oriented basis: whenever a contact between two nodes occurs, a global state is set to denote an open connection between nodes. While the connection is open, the simulator will try to send and receive data. This open state is updated at each step of the simulation until a disconnection occurs. At this moment, any non-finished transfer will be dropped.

Instead, we propose a contact-oriented approach: since we know beforehand the duration of the contacts, we can pre-calculate which messages may be actually transferred along the duration of the connection. This allows us to optimize the number of events we will generate, while keeping the same behavior, instead of keeping the state of the connection at each step of the emulation.

\subsection{Opportunistic Emulation Requirements} % (fold)
\label{sub:opportunistic_network_emulation}

In order to help developers to design, develop and test opportunistic applications, we need to provide a tool that can be easily integrated into the development process. Developers interact with IDE tools where usually testing and debugging tasks are executed to polish applications. We propose the use of network emulation to fill this gap. In order to accomplish the challenges defined before, we define the following basic requirements for such an emulator:

\begin{itemize}
	\item[E1] \textbf{Real-time emulation:} The applications being tested need to operate at the same time scale as in normal operation, hence the need to attach a non complex real-time emulator. Notice that this differs from a simulator or a testbed in the sense that we focus on the application development process and not on the network characterization. 
	%Check and rewrite if necessary

	\item[E2] \textbf{Contact-oriented emulation:} In order to simplify the view of the system, we think that a contact-oriented emulator must be put in place. Contacts are easy to understand in terms of behavior, while hiding the complexity of nodes' mobility. They also allow to know the contact duration beforehand, thus greatly simplifying routing decisions and buffer management. Since we do not rely on a clear link layer technology to deploy opportunistic applications and based on requirements (C1) and (C2), we can abstract the network and still conceive fully functional applications.

	\item[E3] \textbf{Real-time tuning:} Since the parameters of opportunistic algorithms can be cumbersome, we argue that a real-time parameterization is needed to better understand the impact of changes in the emulated opportunistic networks. This lets developers better plan for changes and adapt their applications accordingly.

	\item[E4] \textbf{Real-time monitoring:} In order to tune parameters, the need to observe network and application behaviors, along with their reaction to changes, is very important. Ideally, this observation should be possible during the experiment, and not only afterwards as with simulators. For that, we argue that any opportunistic network emulator needs a real-time monitoring system. 
	
	\item[E5] \textbf{Transparency:} The application that is being tested should not be aware that it runs over an emulator. This also means that we should ideally be able to run the software unmodified. Indeed, to obtain meaningful insights, we need to make sure that the emulation platform itself does not bias application operation. 
	
	\item[E6] \textbf{Repeatability:} To successfully debug an application, we need to exactly reproduce the conditions for an error to happen. Consequently, an opportunistic emulator has to support exact reproduction of simulated network conditions over several distinct runs if necessary for the end user. This includes routing decisions, message generation and contacts. It also means that the emulator needs to have a deterministic behavior, so any part requiring randomness (like synthetic mobility models) should be left out, for example by using precomputed traces. 

	\item[E7] \textbf{Availability:} Unlike a testbed, the emulator should ideally be able to run on more limited hardware resources like a single computer, just like other IDE tools. It cannot afford either to require the booking of resources (hardware and time) in advance, because developer needs may be unpredictable. %+ booking issues (testbed time slots) ?
\end{itemize}

% subsection requirements_for_an_opportunistic_emulator (end)
%!TEX root = main.tex

\section{Conclusions}
\label{sec:conclusions}

%very challenging to conceive applications working on such networks.
%Indeed, opportunistic networking is a promising alternative to infrastructure-based networks. However, its inherent complex dynamics makes it very challenging to conceive applications working on such networks. 

This paper highlights the gap between network characterization and opportunistic applications development. While opportunistic networking is a promising alternative, its inherent complex dynamics makes application design very challenging. Usually, a phase of network characterization is needed to better understand the challenges we will face later during development. Unfortunately, current tools such as DTN simulators or DTN testbeds do not provide any integration with development. Hence, developers must not only deal with network characterization, but also with its impact on the application. 
%We argued that this gap between network characterization and development should be supported by current development tools. we designed a new hybrid emulation system aimed at helping developers to conceive their opportunistic applications.

In this positioning paper, we advocate the use of emulation to reconcile both aspects. We  address the limitations of current network characterization approaches. This allows us to distill important requirements for an opportunistic network emulator. 

%As future work, we plan to use these requirements to propose an emulation architecture. This architecture will be validated through a detailed performance evaluation, then used in a real case study. We also plan to extend the emulator to support the integration with popular DTN stacks. 

\section{Acknowledgments}
The authors would like to thank Tanguy P\'erennou for his suggestions to improve this work. This research was supported in part by the French Ministry of Defense through a financial support of the Direction G\'en\'erale de l'Armement (DGA).

%{\small 
\bibliographystyle{ieeetr}
\bibliography{stateofart_short}
%}

\end{document}